\documentclass[aps,prl,reprint,superscriptaddress,longbibliography]{revtex4-2}

\usepackage{amsmath}
\usepackage{amssymb}
\usepackage{wasysym}
\usepackage{graphicx}
\usepackage{hyperref}
\usepackage{xcolor}
\usepackage{bm}  
\usepackage{physics2}
\usepackage{orcidlink}
\usephysicsmodule{ab, ab.braket}  
\usepackage{siunitx}  
\usepackage{csquotes}
\usepackage{soul}

\begin{document}

\renewcommand{\tableautorefname}{Tab.}
\renewcommand{\figureautorefname}{Fig.}
\renewcommand{\equationautorefname}{Eq.}

\title{Collective cluster nucleation dynamics in quantum magnets }

\author{Philip~Osterholz${}^{\orcidlink{0009-0007-3444-3183}}$}
\author{Fabio~Bensch${}^{\orcidlink{0009-0000-1051-3297}}$}
\author{Shuanghong~Tang${}^{\orcidlink{0000-0002-1878-0311}}$}
\author{Silpa~Baburaj~Sheela${}^{\orcidlink{0000-0003-4415-6288}}$}
\affiliation{Physikalisches Institut, Eberhard Karls Universit\"at T\"ubingen, Auf der Morgenstelle 14, 72076 T\"ubingen, Germany}
\author{Björn~Sbierski${}^{\orcidlink{0000-0003-1063-2389}}$}
\affiliation{Institut f\"ur Theoretische Physik, Universit\"at T\"ubingen, Auf der Morgenstelle 14, 72076 T\"ubingen, Germany}
\author{Igor~Lesanovsky${}^{\orcidlink{0000-0001-9660-9467}}$}
\affiliation{Institut f\"ur Theoretische Physik, Universit\"at T\"ubingen, Auf der Morgenstelle 14, 72076 T\"ubingen, Germany}
\affiliation{Center for Integrated Quantum Science and Technology, Universit\"at T\"ubingen, Auf der Morgenstelle 14, 72076 T\"ubingen, Germany}
\author{Christian~Gro\ss${}^{\orcidlink{0000-0003-2292-5234}}$}
\email{christian.gross@uni-tuebingen.de}
\affiliation{Physikalisches Institut, Eberhard Karls Universit\"at T\"ubingen, Auf der Morgenstelle 14, 72076 T\"ubingen, Germany}
\affiliation{Center for Integrated Quantum Science and Technology, Universit\"at T\"ubingen, Auf der Morgenstelle 14, 72076 T\"ubingen, Germany}

\date{\today}

\begin{abstract}
	Strongly interacting many-body systems exhibit collective properties that emerge from complex correlations among microscopic degrees of freedom.
	These cooperative phenomena govern the non-equilibrium response of quantum systems, with relevance ranging from condensed matter physics to quantum field theories describing fundamental aspects of our universe~\cite{coleman1977-cg, turner1982-cg, kormos2017-cg, yin2025-cg}.
	Understanding such emergent dynamics from first principles remains one of the central challenges in quantum many-body physics.
	Here we report on the observation of collective cluster nucleation dynamics following quenches in 2D ferromagnetic quantum Ising systems implemented in an atomic Rydberg array~\cite{browaeys2020-cg}.
	Our experiments reveal two distinct regimes:
	In the confined regime, we observe an energy-dependent cluster size, revealing large collective clusters exceeding ten spins.
	In contrast, the deconfined regime is characterized by kinetically constrained, avalanche-like nucleation dynamics involving the entire system.
	Our findings establish a new frontier for quantum simulations with Rydberg arrays, enabling controlled exploration of non-equilibrium phenomena previously out of reach.
	Beyond advancing experimental capabilities, they provide fundamental insights into highly correlated processes with implications that reach from quantum magnetism and glassy dynamics to cosmological models of the early universe~\cite{biroli2013, balducci2022-cg, pavesic2025-cga, lagnese2024-cg}.
\end{abstract}

\maketitle

Correlated dynamics underlie a wide range of collective phenomena, including metastability and false vacuum decay~\cite{rutkevich1999-cg, lagnese2021-cg, yin2025-cg}, bubble nucleation and confinement~\cite{kormos2017-cg, gribben2018-cg, liu2019-cg, pavesic2025-cg}, and kinetically constrained dynamics with subsequent slow thermalization due to Hilbert space fragmentation~\cite{everest2016-cg, balducci2022-cg, hart2022-cg, darbha2025-cg}.
The quantum Ising model in transverse and longitudinal fields provides a minimal yet versatile model for exploring many aspects of such non-equilibrium physics.
With only a few parameters, it captures the interplay among interactions, external fields, and quantum or thermal fluctuations~\cite{suzuki2012-cga}.
Experimentally, the model has been realized in atomic, ionic, and solid-state quantum simulators, and many aspects of its dynamics have already been explored~\cite{browaeys2020-cg, blatt2012, mi2022-cg}.
In one-dimensional lattices, metastability and domain-wall confinement have been observed, arising from a size-dependent domain energy~\cite{tan2021-cg, zhu2024-cg, luo2025-cg, vodeb2025-cg}.
Closely related is the phenomenology of false vacuum decay, namely, the collective transition from one ordered quantum state to another.
This has been investigated in the continuum using Bose-Einstein condensates in effectively one-dimensional settings~\cite{zenesini2024-cg}.
In higher dimensions, the physics becomes substantially richer.
Kinetically constrained magnetic dynamics emerge, where the spatial shape of collective spin domains becomes important~\cite{balducci2022-cg, hart2022-cg, pavesic2025-cg, kaltenmark2025-cg}.
Recently, spectral signatures of small confined clusters have been observed in two-dimensional Rydberg arrays~\cite{darbha2025-cg} and
string-breaking, a paradigmatic manifestation of confinement in gauge theories, has been experimentally studied~\cite{gonzalez-cuadra2025-cg}.
At the so-called deconfinement point, the energy cost of growing an existing cluster by flipping an adjacent spin is balanced by the energy gain from the external field.
Avalanche-like domain growth under strong geometric constraints is predicted in this regime~\cite{everest2016-cg}.
Analogous avalanche dynamics associated with first-order quantum phase transitions have been explored both in optical lattices~\cite{song2022-cg} and in atomic systems with competing short- and long-range interactions in optical cavities~\cite{hruby2018-cg}.

In our experiments, we study collective cluster formation in two-dimensional square-lattice Ising systems, realized in atomic arrays with strong van der Waals interactions between Rydberg atoms.
While the majority of experiments on this platform use the Rydberg blockade for implementing many-body spin models~\cite{browaeys2020-cg}, we rely on an anti-blockade, in which Rydberg atoms facilitate the excitation of nearby atoms~\cite{ates2007-cg}.
In this parameter regime, spin-motion coupling demands exquisite control over the atomic positions to implement unitary dynamics within the many-atom pseudo spin-1/2 subspace formed by the atomic ground and one Rydberg state~\cite{li2013-cg, marcuzzi2017-cg, schlegel2025-cg}.
Indeed, coherent dynamics have recently been demonstrated, and quantum many-body scars and kinetically constrained dynamics in one-dimensional arrays have been studied~\cite{zhao2025-cg, datla2025-cg}.
Earlier experiments explored Rydberg facilitation in continuum systems, where dissipation played a major role~\cite{schempp2014-cg,malossi2014-cg, urvoy2015-cg, helmrich2020-cg, brady2023-cg}.
In contrast, our system is initially at zero temperature in the pseudo-spin sector, and dissipation has little effect on the fast timescales on which we study the collective dynamics.
With the experiments reported here, we demonstrate that facilitated Rydberg dynamics can be coherently controlled in large two-dimensional arrays.
This paves the way to study a wide range of new phenomena on these versatile quantum simulators, including exotic equilibrium quantum states~\cite{zhang2024-cg}, dynamics of collective spin-clusters~\cite{pavesic2025-cga}, thermalization in strongly constrained systems~\cite{everest2016-cg, balducci2022-cg}, and equilibration in glassy systems~\cite{pancotti2020-cg}.

\begin{figure}[t]
	\centering
	\includegraphics{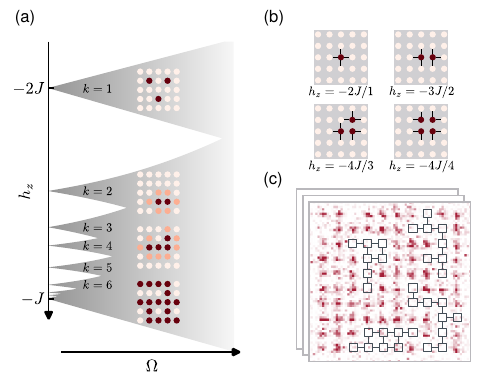}
	\caption{
		\textbf{Illustration of the many-body spectrum in a 2D Ising model.}
		\textbf{(a)} Sketch of the quench response of the square-lattice Ising model with nearest-neighbor interactions for an initial state with all spins in the $\ket|\downarrow>$-state (light circles) as a function of $h_z$ and $\Omega$.
		For quenches to weak transverse field $\Omega$, the response is sharp and concentrated around discrete many-body resonances in $h_z$.
		At $h_z=h_z^S=-2J$, single spin flips (dark red circles) are resonant, while confined clusters of increasing cluster size $k$ become resonant at larger $h_z$ up to the deconfinement point at $h_z=h_z^A=-J$.
		Here, the dynamics are characterized by avalanche-like growth of clusters.
		With increasing $\Omega$, the resonances broaden (indicated by the gray shading), shift, and the cluster sizes mix due to quantum fluctuations (light red circles).
		\textbf{(b)} Confined clusters behave as collective objects with characteristic shapes determined by energetic constraints.
		In the classical ($\Omega \rightarrow 0$) limit, the excitation energy of a cluster is determined by the number of opposite-spin bonds (black lines) along its perimeter.
		The collective resonance is located at the total cluster energy divided by the number of spins in the cluster, as indicated below the cluster sketches.
		The square-shaped cluster at the lower right is located at an energy of $h_z=-J$, where the addition of further spins is resonant such that the cluster size is not confined.
		\textbf{(c)} Typical experimental snapshot after a quench to the deconfinement resonance at $h_z = -J$.
		Four clusters of Rydberg atoms, that is, connected sites of flipped spins identified by missing atoms, are highlighted by the linked black squares.
	}
	\label{fig:1}
\end{figure}

\begin{figure*}[t]
	\centering
	\includegraphics{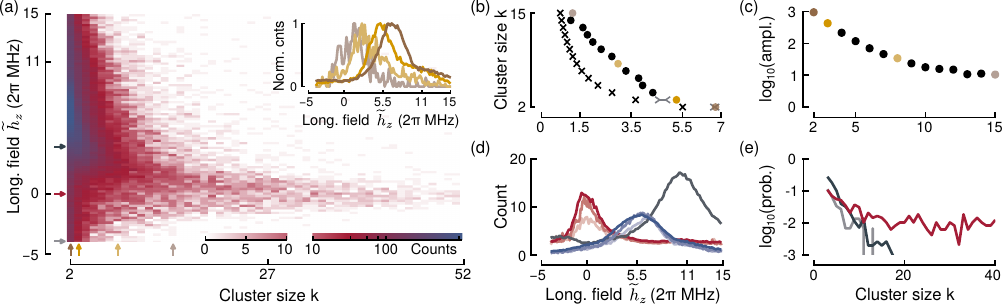}
	\caption{
		\textbf{Spectral response.}
		\textbf{(a)} Two-dimensional histogram of the cluster size distribution versus longitudinal field after $\qty{2}{\micro \second}$ evolution time with $\Omega=2\pi \times \qty{2.24}{\mega \hertz}$.
		To highlight the tails of the distribution, we use a linear color scale up to 10 counts and a logarithmic scale above.
		The inset shows normalized cuts along the vertical axis for 2-, 3-, 8-, and 15-atom clusters as indicated by the colored arrows on the bottom.
		\textbf{(b)} Resonance positions extracted from Gaussian fits to the data shown in the inset of (a) versus cluster size (dots) together with the classical expectation (black crosses).
		The colored dots correspond to the cuts shown in (a).
		The gray markers for $k=2,3$ indicate resonance positions as predicted by second-order perturbation theory~\cite{supplement}.
		\textbf{(c)} Amplitude of the Gaussian fits versus cluster size.
		The coloring is as in (b).
		\textbf{(d)} Mean count of isolated flipped spins (gray) and mean count of clusters (blueish colors) together with the mean $\uparrow$-spin count in clusters, respectively cluster size (reddish colors), versus longitudinal field.
		The lighter colors for the cluster number and size indicate data taken at shorter evolution times of $\qty{0.5}{\micro \second}$ (light) and $\qty{1}{\micro \second}$ (medium-light).
		\textbf{(e)} Normalized cluster size distribution (horizontal cuts through the data shown in (a) as marked by the arrows on the left) at the facilitation resonance (red) and at $\widetilde{h}_z = \pm 2\pi \times \qty{4}{\mega \hertz}$ (dark, light gray)  after $\qty{2}{\micro \second}$ evolution time.
	}
	\label{fig:2}
\end{figure*}

Our experiments are based on potassium-39 atoms in optical tweezer arrays~\cite{lorenz2021-cg}.
We start with the preparation of a square-lattice array of $15\times15$ tweezers with a nearest-neighbor distance of $d_0=\qty{7}{\micro\meter}$, and load a single atom in each tweezer with about $\qty{65}{\percent}$ probability.
This loading is done from a reservoir light sheet and under continuous gray molasses cooling~\cite{supplement}.
We then arrange the atoms into a central $10\times10$ array of $N=100$ atoms and assure its isolation by clearing atoms from the tweezers surrounding the array.
In a single sorting step, we reach an occupation of $\qty{92}{\percent}$ in the target region.
The tweezer array is carefully position corrected and homogenized to sub-percent peak-to-peak intensity deviations~\cite{chew2024-cg}, an essential step for high-fidelity imaging in our tweezers formed by $\qty{1064}{\nano\meter}$ light and for improving Raman sideband cooling.
The latter we perform in three dimensions, and we reach a ground-state population of $\qty{90}{\percent}$ in the radial (in-plane) tweezer directions and of $\qty{70}{\percent}$ in all directions.
This is critical to mitigate position-fluctuation-induced disorder.
The final step to prepare the initial state is to optically pump the atoms to the 4S$_{1/2}$ $\ket|F=2,m_F=2> \equiv \ket|\downarrow>$ hyperfine state.

We couple this state to the 75S $\ket|J=1/2, m_J=1/2> \equiv \ket|\uparrow>$ Rydberg state using a two-photon transition via the intermediate 5P$_{3/2}$ $\ket|F=3/2,m_F=3/2>$ state.
We reach maximal Rabi frequencies $\Omega \approx 2\pi\times\qty{2}{MHz}$ across the entire central array.
The interaction results in an energy shift $U_0 = 2\pi\times \qty{11}{\mega\hertz}$ for two Rydberg atoms at the lattice distance $d_0$.
Repulsive forces between Rydberg atoms limit the observation time and constrain the maximal controllable interaction strength~\cite{supplement}.
The resulting antiferromagnetic Ising Hamiltonian is
\begin{equation}
	\begin{aligned}
		\hat{H}/\hbar & = \sum_{i \neq j} \frac{U_{ij}}{2} \hat{R}_i\hat{R}_j - \frac{\Delta}{2} \sum_i \hat{Z}_i + \frac{\Omega}{2} \sum_i \hat{X}_i       \\
		              & \approx \frac{J}{4} \sum_{<i,j>} \hat{Z}_i\hat{Z}_j - \frac{h_z}{2} \sum_i \hat{Z}_i + \frac{\Omega}{2} \sum_i \hat{X}_i, \nonumber
	\end{aligned}
\end{equation}
with $\hbar$ being the reduced Planck constant and $U_{ij}=U_0/r_{ij}^6$ is the van der Waals interaction between atoms $i,j$ spaced by $r_{ij}d_0$.
We have written the Hamiltonian in standard Ising form in terms of the Pauli operators $\hat{X}_i$, $\hat{Z}_i$ at site $i$ by using the relation $\hat{R}_i =(\hat{Z}_i + \hat{\mathbb{I}}_i)/2$ with the local projector to the Rydberg state $\hat{R}_i$ and the identity operator $\hat{\mathbb{I}}_i$.
The interaction strength between nearest neighbors is $J = U_0$ and the laser detuning $\Delta$ controls the longitudinal field $h_z = \Delta-\mathcal{N} U_0/2$, with the number of nearest neighbors $\mathcal{N}=4$.
In rewriting the Hamiltonian of the second line, we neglected beyond nearest-neighbor interactions of strength $U_0/8$ or lower.
While these interactions increase the cluster energies quantitatively, they do not change the qualitative physics, and we will comment on their effect where appropriate.
We also neglected local longitudinal fields for sites with fewer neighbors emerging on the edge of the system and around vacant sites.

The high-energy spectrum of our antiferromagnetic case maps to the low-energy spectrum of the ferromagnetic Ising model, in which cluster formation and metastability are usually discussed~\cite{kormos2017-cg, lagnese2021-cg, darbha2024-cg}.
In the classical limit $\Omega=0$ and for $h_z=0$, the fully polarized states are degenerate and the highest energy states of the Hamiltonian.
For a general state, the number of bonds connecting spins of opposite orientation, which constitute domain walls, determines the energy difference to the two extremal states.
A finite longitudinal field $h_z$ biases one of the spin orientations against the other, such that the domain wall energy cost can be compensated.
For the system being initially in the $\ket|\downarrow>^{\otimes N}$ state, this leads to a characteristic spectrum of resonances in $h_z$, where confined clusters of $k$ flipped spins and a given domain wall length become degenerate with the initial state.
The single spin flip resonance is at an energy of $h_z^S=-2J$, and at higher energies resonances of larger cluster sizes $k$ follow (see \autoref{fig:1}a).
The energetic position of the collective resonances is determined by the equality of the surface energy gain ($-J/2$ per opposite-spin bond) to the volume energy cost ($h_z$ per flipped spin), resulting in $h_z^{C}(k) = - J(k+1)/k$.
This series of many-body resonances is constrained to feature no loops in the confined cluster shapes.
The energetic cascade of $k$-sized clusters is bounded from above by a special resonance at $h_z^{A} = -J$.
At this deconfinement point, all clusters featuring a single loop are resonant, independent of their size.
Additionally, clusters with a certain number of kinks (see \autoref{fig:1}b, lower-left) are in or near resonance due to the contribution of diagonal interactions, which increase the energy of such clusters.
A finite transverse field $\Omega$ induces quantum fluctuations leading to a broadening of the resonances and to a renormalization of the confined cluster energies $h_z^{C}(k)$~\cite{supplement}.

In a first set of experiments, we aim to explore this collective spectral response.
We start in the fully polarized $\ket|\downarrow>^{\otimes N}$ state.
For different $h_z$, we abruptly quench the transversal coupling from zero to $\Omega = 2\pi\times\qty{2.24}{MHz}$ and take snapshots of the system after an evolution time of $\qty{2}{\micro \second}$.
Spin-flipped atoms, that is, atoms excited to the Rydberg state, are lost from the traps due to the anti-trapping nature of the Rydberg state in our optical tweezers.
We reconstruct the spin configuration in the $\hat{Z}$-basis by comparing the tweezer population before and after the quench.
\autoref{fig:1}c shows a typical final image where we highlight connected clusters of Rydberg atoms.
By repeating the experiment for different $h_z$, we obtain statistics of the cluster size distribution.
In the following, we measure the longitudinal field $\widetilde{h}_z = -(h_z - h_z^A)$ inverted and relative to the deconfinement resonance.
The resulting two-dimensional histogram is shown in \autoref{fig:2}a, where we include only clusters, defined as two or more adjacent spin flips.
Due to the Rydberg blockade, no feature is visible at the single spin flip resonance $\widetilde{h}_z = \widetilde{h}_z^S = 2\pi \times \qty{11}{\mega \hertz}$.
For decreasing $\widetilde{h}_z$, the cluster size distribution develops a pronounced tail towards large clusters.
Vertical cuts at different cluster sizes (inset of \autoref{fig:2}a) reveal a series of resonances, which show the expected trend of a smaller $\widetilde{h}_z$ for larger cluster size.

In \autoref{fig:2}b, we summarize the resonance positions extracted from Gaussian fits to the cuts together with the classical expectation $\widetilde{h}_z^{C}(k) = J/k$.
Our analysis reveals collective resonances up to confined cluster sizes of about $k=15$.
We observe a sizable shift of the collective cluster resonances towards larger $\widetilde{h}_z$, which we attribute to the quantum fluctuations induced by the transverse field $\Omega$.
This interpretation is supported by second-order perturbation theory in $\Omega/U_0$.
Explicit calculations for $k=2$ show remarkable agreement with the measured value.
For larger clusters, quantum fluctuations as well as the neglected longer-range interactions lead to shape-dependent energies.
The longer-range interactions are expected to shift the resonances towards smaller $\widetilde{h}_z$, opposite to the shifts due to quantum fluctuations.
In the analysis of the experimental data, we do not resolve the cluster shape, but we calculated the range spanned by the resonance frequencies for the different cluster shapes theoretically.
The experimentally measured values fall outside this range as displayed for $k=3$ in \autoref{fig:2}b and detailed in~\cite{supplement}.
We attribute this effect to the increasing importance of quantum fluctuations with cluster size, which require higher-order perturbation to be considered.

The amplitudes of the resonances are plotted in \autoref{fig:2}c.
We observe a clear deviation from an exponential suppression in the coupling to larger clusters, which would be expected from the increasing number of intermediate off-resonant states involved in the coupling.
We attribute this deviation to a strongly growing number of coupling terms with increasing cluster size.
\autoref{fig:2}d highlights the existence of three qualitatively different regimes in the spectrum.
The number of isolated flipped spins peaks around the single spin-flip resonance, while smaller clusters dominate the response for intermediate $\widetilde{h}_z$.
Finally, on the deconfinement resonance, the size of the clusters peaks, and the number of clusters is minimal.
The data in \autoref{fig:2}d also shows that the system requires longer times to reach a steady state on the deconfinement resonance and in the regime where large clusters form.
The cluster size distribution changes qualitatively on the deconfinement point.
While it is strongly peaked in the confined regime, it becomes broad on resonance (see \autoref{fig:2}e).

\begin{figure}[t]
	\centering
	\includegraphics{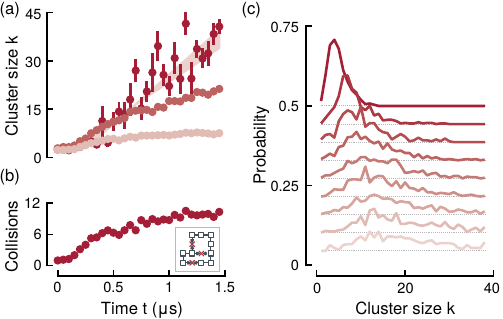}
	\caption{
		\textbf{Deconfined cluster formation kinetics.}
		All data is taken on the deconfinement resonance at $\widetilde{h}_z = \widetilde{h}_z^A = 0$ and with $\Omega = 2\pi\times\qty{2.0}{MHz}$.
		\textbf{(a)} Growth dynamics for the two largest clusters in the system.
		For each run, we identify the two largest clusters and average their sizes individually over all runs (middle red for the largest, light red for the second-largest cluster).
		In dark red, we show the mean size of the largest cluster conditioned to runs in which only one cluster was identified.
		A linear fit reveals a growth rate of $\qty{24 \pm 4}{\text{sites} \per \micro \second}$, where the uncertainty is dominated by the selection of data points included in the fit (all vs. $t>\qty{0.25}{\micro \second}$).
		Error bars indicate the standard error of the mean.
		\textbf{(b)} Number of cluster collisions versus time.
		A collision is defined as a bond on which cluster-growth is blocked by a nearby cluster, that is, where we identify two adjacent $\uparrow$-spins around a $\downarrow$-spin and where both $\uparrow$-spins belong to distinct clusters (see inset).
		\textbf{(c)} Histograms showing the evolution of the distribution of cluster sizes.
		We bin the data in intervals of $\qty{0.15}{\micro \second}$ and show the area-normalized counts for increasing evolution times from
		$\qty{0.1}{\micro \second}$ to $\qty{1.45}{\micro \second}$ (dark to light).
		The individual histograms are offset vertically for better visibility, as indicated by the thin gray lines.
	}
	\label{fig:3}
\end{figure}

We now turn our focus to the study of the kinetics of the formation of deconfined clusters.
To increase the probability of observing cluster-growth, we coherently increase the population in $\ket|\uparrow>$ initially before quenching to the deconfinement resonance.
Specifically, we pulse the Rydberg laser resonantly for a short time to prepare about $\qty{4}{\percent}$ of the atoms delocalized in the $\ket|\uparrow>$ state.
While the transverse field is zero for $\qty{200}{\nano\second}$, we jump $h_z$ to the deconfinement resonance and then switch on the transverse field with $\Omega=2\pi\times\qty{2.0}{MHz}$.
In \autoref{fig:3}a we show the mean size of the two largest clusters as a function of time together with the mean size of the cluster conditioned to only one cluster identified in the run.
Discarding the short-time dynamics up to about $\qty{250}{\nano \second}$ which is influenced by the finite rise time of our laser pulses~\cite{supplement}, we observe rapid and approximately linear-in-time growth of the clusters.
In the single cluster case, we extract a constant rate $R = (12\pm2\,)\Omega$.
This rate is much larger than expected from the maximum group velocity for domain growth in 1D $R_{1D} = 2\Omega$, which follows analytically from the motion of free domain walls~\cite{fogedby1978-cg, kormos2017-cg}.
We attribute this speed-up to the presence of multiple bonds, over which growth is active simultaneously.
If several clusters are present, the rate is initially unchanged, but the growth slows down for the largest cluster at later times, when the size of the second largest cluster approximately saturates.
We interpret this simultaneous slowdown as a signature of cluster-cluster interactions, which reduce the number of bonds available for growth.
This picture is supported by the data in \autoref{fig:3}b, where we show the number of cluster collisions, defined as the number of bonds on which cluster-growth is blocked by a nearby cluster.
The collisions show qualitatively similar dynamics as the mean largest cluster size, indicating signs of saturation when the cluster-growth slows down.
Finally, we show the temporal evolution of the distribution of cluster sizes in \autoref{fig:3}c.
From being initially peaked, the distribution broadens quicker than its mean value shifts towards larger sizes.
The effect of the initially increased Rydberg population is to suppress small clusters, as reflected in the difference of the late time distribution to the steady state histogram obtained without the initial pulse (cf. \autoref{fig:2}e).

\begin{figure}[t]
	\centering
	\includegraphics{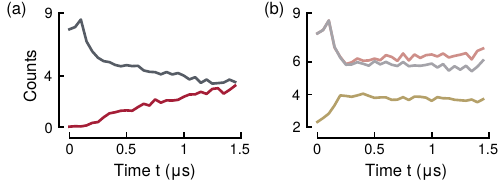}
	\caption{
		\textbf{Shape constraints and cluster number conservation.}
		\textbf{(a)} The number of detected loops in the clusters versus time (red) and the number of isolated flipped spins (gray) show a similar increase (decrease) rate after an initial transient behavior.
		\textbf{(b)} The light red data shows the sum of the detected loops and isolated flipped spins.
		For times larger than $\qty{0.25}{\micro \second}$, this sum is approximately constant.
		The gray line shows the sum of isolated flipped spins and the number of spins in clusters, which need to be flipped to remove all loops.
		The yellow data corresponds to the mean number of detected clusters.
	}
	\label{fig:4}
\end{figure}

To shed light on the constraints governing the shape of the deconfined clusters, we analyze the number of loops within clusters as a function of time.
This number is expected to remain constant, as closing a loop means to increase the cluster perimeter by less than two opposite-spin bonds; thus, this process is off-resonant.
In \autoref{fig:4}a we show the evolution of the number of detected loops and of isolated $\ket|\uparrow>$-spins.
After an initial phase, in which clusters have not grown substantially, we observe a slow linear growth of the number of detected loops.
Simultaneously, the number of isolated $\ket|\uparrow>$-spins decreases at a similar rate.
We attribute this slow trend to detection errors due to our finite recapture fidelity (ca. \qty{95}{\percent}).
As clusters grow with time, the probability for such errors to appear as isolated spin-flips reduces, while the probability to cause falsely detected loops increases.
This interpretation is supported by \autoref{fig:4}b, where we show the sum of the two quantities.
To account for cases in which a single detection error increases the loop count by more than one, we refine the analysis further.
For each individual snapshot, we determine the minimal number of detection errors required to account for all observed loops.
Adding this quantity to the number of detected  isolated spin-flips yields a fully constant signal.
Hence, the number of loops is indeed conserved on the probed timescale.
The same kinetic constraints that enforce loop-number conservation also prevent clusters from merging during the dynamics.
In \autoref{fig:4}b we also show the evolution of the detected mean cluster number, which indeed saturates quickly and then slightly decreases, which we also attribute to our finite detection fidelity.

In conclusion, we have presented experimental studies of confined and deconfined cluster formation after quenches in the 2D transverse field Ising model.
Our data reveals an intriguing collective response of this paradigmatic quantum many-body system.
We observed large confined clusters with resonances, shifted by quantum fluctuations.
On the deconfinement resonance, we observe fast avalanche-like growth of clusters with strong constraints on the cluster shapes and strong mutual interactions.
We presented signatures of a slowdown of the dynamics at later times.
Reaching the steady state is hindered by the strong interactions between the extended clusters, effectively constraining each others growth, a characteristic feature of glassy dynamics.
Our work opens several exciting perspectives for future studies of metastability and confinement in 2D lattice systems~\cite{pavesic2025-cg, pavesic2025-cga}, among them the functional dependence of the confined cluster energies and coupling strengths on the transverse field, false vacuum decay dynamics, or the role of long-range interactions~\cite{kaltenmark2025-cg}.
Changing the 2D array geometry to a Kagome lattice, will allow us to explore lattice gauge theories~\cite{homeier2025-cg}.
Furthermore, our experiments demonstrate coherent facilitated Rydberg dynamics in two-dimensional arrays, opening a new research line with Rydberg arrays and paving the way towards further studies of the physics of Rydberg quantum magnets under facilitation constraints~\cite{everest2016-cg, balducci2022-cg, hart2022-cg}.

\bigskip
\begin{acknowledgments}
	\textbf{Acknowledgments:} We acknowledge discussions with Hannes Pichler, Marco di Liberto, Simone Montangero, Federica Surace and Johannes Zeiher.
	This work received funding from the Horizon Europe program HORIZON-CL4-2022-QUANTUM-02-SGA via the project 101113690 (PASQuanS2.1) and via the Horizon-MSCA-Doctoral Network QLUSTER (HORIZONMSCA-2021-DN-01-GA101072964), the Federal Ministry of Education and Research Germany (BMBF) via the project 13N15974, and the Deutsche Forschungsgemeinschaft within the research units FOR5413 (Grant No. 465199066), and through FOR 5522 (Grant No. 499180199) and the SPP 1929 GiRyd (project GR4741/5).
	We also acknowledge funding through JST-DFG 2024: Japanese-German Joint Call for Proposals on “Quantum Technologies” (Japan-JST-DFG-ASPIRE 2024) under DFG Grant No. 554561799 and from the Alfried Krupp von Bohlen and Halbach Foundation. IL acknowledges support from the European Union through the ERC grant OPEN-2QS (Grant No. 101164443).
\end{acknowledgments}

\bigskip
\textbf{Author Contributions:} All authors contributed extensively to the
planning, data acquisition, or analysis of the results presented here.

\bigskip
\textbf{Data availability:}
The experimental and theoretical data and evaluation scripts that support the findings of this study will be available on Zenodo.

\bigskip
\textbf{Competing interests:}
There are no competing interests to declare.

\bigskip
\textbf{Note added:}
During the preparation of this manuscript, we became aware of a related study in a 1D Rydberg array by the group of Prof. Li You at Tsinghua University Beijing.

\newpage
\clearpage

\renewcommand{\thefigure}{S\arabic{figure}}
\renewcommand{\thetable}{S\arabic{table}}
\renewcommand{\theequation}{S\arabic{equation}}
\renewcommand{\thepage}{S\arabic{page}}

\onecolumngrid

\section*{Supplemental Information:}

\section*{Tweezer array and high fidelity detection}

The 225 optical tweezers used in this experiment are generated from a $\qty{1064}{\nano\meter}$ single-mode fiber laser and positioned using a liquid-crystal spatial-light modulator placed in the Fourier plane of our in-vacuum objective.
The objective has a numerical aperture of 0.6 and is covered with a thin gold mesh that isolates the atoms from stray electric fields that may originate from its non-conducting surface.

We detect single $^{39}$K atoms in the tweezers by inducing fluorescence on the D2 line while simultaneously cooling the atoms inside the traps in a gray-molasses configuration using the D1 line for $\qty{100}{\milli\second}$~\cite{angonga2022-cg}.
To enhance the survival probability to $\qty{99}{\percent}$, we pulse the D2-light with a frequency of $\qty{1}{\kilo\hertz}$.
The scattered photons from the D2-light are spectrally separated from the D1-light by four consecutive band-pass filters and imaged onto an EMCCD camera.
While the $4S_{1/2}$ ground state has a comparably small polarizability at $\qty{1064}{\nano\meter}$ and experiences negligible vector light shifts, the $4P_{1/2}$ and $4P_{3/2}$ excited states of the D1 and D2 line exhibit light shifts that are more than six times larger, along with sizable vector components.
Due to the strong sensitivity of our cooling and imaging scheme to light-shifts of the transitions, this makes our imaging a sensitive probe of the tweezers depths.
To ensure high-fidelity detection across the entire tweezer array, we homogenize the intensities of the tweezers following the procedure reported in ref.~\cite{chew2024-cg}.
The local feedback signals for the individual tweezers are extracted from the measured fluorescence strengths of the tweezers.
We typically reach a homogeneity of around $\qty{1}{\percent}$ limited by statistical fluctuations.

\section*{Initial state preparation}

After an initial MOT loading and compression phase, we cool into our static tweezer array using a gray molasses.
To facilitate the loading of the tweezers, we simultaneously apply a strongly elliptical dipole trap with a $\qty{7}{\micro\meter}$ ($\qty{130}{\micro\meter}$) vertical (horizontal) waist.
This light-sheet trap enters the vacuum chamber orthogonal to the tweezer beam direction and is aligned to overlap with the focal plane tweezer array.
We use $\qty{8}{\watt}$ of $\qty{1064}{\nano\meter}$ laser light to form the light sheet.
Note that the Rayleigh range is only about $\qty{150}{\micro\meter}$, providing also confinement in the propagation direction.
We empirically found that the light-sheet trap helps to stabilize the loading probability of the tweezers against power drifts in the cooling beams.
It also increases the confinement along the weaker trapping axis of the tweezer array.
The same beam configuration is used during parity-projection by the gray molasses beams, after which only single atoms remain in the tweezers~\cite{lester2015-cg}.
We achieve a stable loading of $\qty{65}{\percent}$ single atom probability per tweezer for a typical loading time of $\qty{400}{\milli\second}$.

Subsequently, we take a first picture to determine which optical tweezers were loaded.
We then use a crossed acousto-optical deflector (AOD) to reshuffle the atoms for high filling in the central $10 \times 10$ tweezers.
The AOD tweezer is generated using a $\qty{795}{\nano\meter}$ DBR diode, providing $\qty{80}{\milli\watt}$.
In a single sorting run, we achieve a filling fraction in the central array of $\qty{92}{\percent}$.
The same mobile tweezer is used to remove the atoms from all sites neighboring the central $10 \times 10$ array.

\begin{figure}[t]
	\centering
	\includegraphics[]{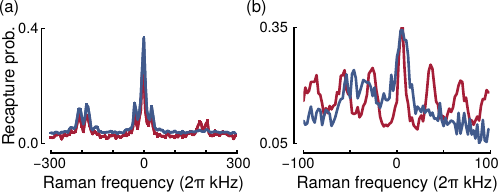}
	\caption{\textbf{Raman sideband cooling.}
		In red we show the Raman spectra without sideband cooling and in blue with sideband cooling applied.
		\textbf{(a)} Radial Raman spectrum.
		\textbf{(b)} Axial Raman spectrum.
	}
	\label{fig:si1}
\end{figure}

To prepare the atoms close to the motional ground state of the trap, we apply Raman sideband cooling in the tweezer array and reach a radial (axial) ground-state population of $\qty{90}{\percent}$ ($\qty{85}{\percent}$).
\autoref{fig:si1} shows radial and axial Raman spectra with and without Raman sideband cooling.
This step is critical to avoid positional disorder that strongly affects the facilitation dynamics.
The $\qty{1064}{\nano\meter}$ tweezers induce strong tensor, and for elliptical light also strong vector, light shifts of the $4\mathrm{P}$ state manifold.
Of particular importance is the $4\mathrm{P}_{1/2}$ state used for repumping in the Raman cooling process~\cite{lorenz2021-cg}.
The performance of the cooling critically depends on the darkness of the motional ground state, and we found that the state mixing associated with vector light shifts in the excited state quickly deteriorates this darkness.
This effect is minimized, by choosing linear polarization of the tweezer light to avoid strong vector light shifts.
For the last cycle of the Raman sideband cooling process, we keep the repumping beam on longer to reach a $\qty{99}{\percent}$ population in the $4\mathrm{S}_{1/2}$ $\ket|F=2, m_F=2>$ state.
This sequence initializes the system to near the motional ground state and in the paramagnetic state $\ket|\downarrow>^{\otimes N}$.

\section*{Rydberg coupling}

We induce the transverse and longitudinal fields by laser-coupling the ground state to selected Rydberg states.
From the corresponding dressed-state picture, it becomes evident that the Rabi frequency $\Omega$ maps directly to the transverse field and the detuning $\Delta$ to the longitudinal via $h_z = (\Delta - \frac{\mathcal{N} U_0}{2})$, with $U_0$ being the interaction shift obtained from the pair interaction potential of two Rydberg atoms and $\mathcal{N}$ the number of nearest neighbors.
The transverse field is implemented by two-photon laser coupling to the Rydberg state.
Our ladder scheme connects the $4\mathrm{S}_{1/2}$ $\ket|F=2,m_F=2>$ near-resonantly, with a detuning of about $2\pi \times \qty{500}{\mega\hertz}$ to the intermediate state $5\mathrm{P}_{3/2}$ \(\ket|F=3,m_F=3>\), and then to the $75\mathrm{S}$ $\ket|J=1/2,m_J=1/2>$ Rydberg state, while applying $5\,\text{G}$ offset field along the beams propagation axis.

The transition from $4\mathrm{S}_{1/2}$ $\ket|F=2,m_F=2>$ to $5\mathrm{P}_{3/2}$ $\ket|F=3,m_F=3>$ is driven with a titanium–sapphire laser which is doubled to $\qty{405}{\nano\meter}$, and about $\qty{50}{\milli\watt}$ is guided to the atoms.
To couple from the $5\mathrm{P}_{3/2}$ $\ket|F=3,m_F=3>$ to the $75\mathrm{S}$ $\ket|J=1/2,m_J=1/2>$ state, we use another titanium–sapphire laser at $\qty{974}{\nano\meter}$, from which $\qty{180}{\milli\watt}$ reach the atoms.
Both lasers are stabilized to resonators made from ultralow-expansion ceramics using Pound–Drever–Hall schemes and feature state-of-the-art phase-noise performance with spectra similar to those measured in refs.~\cite{jiang2023-cg, denecker2025-cg}.
Both beams pass acousto-optical modulators for precise control over the pulses reaching the atoms.
To enhance reproducibility of pulse amplitudes, we actively stabilize the laser intensities using photodiodes in a sample-and-hold fashion.

In the experiments described in the main text fast switching of the Rydberg coupling is important.
This is achieved by switching on the coupling from $5\mathrm{P}_{3/2}$ $\ket|F=3,m_F=3>$ to the $75\mathrm{S}$ $\ket|J=1/2,m_J=1/2>$ state shortly before the coupling from $4\mathrm{S}_{1/2}$ $\ket|F=2,m_F=2>$ to $5\mathrm{P}_{3/2}$ $\ket|F=3,m_F=3>$ state.
The switching time is limited by the double pass AOM in the $\qty{405}{\nano\meter}$ beam path, which is about $\qty{200}{\nano\second}$.
The effective detuning to the Rydberg state is also controlled by varying the AOM frequency in the $\qty{405}{nm}$ beam path, while the frequency applied to the AOM in the $\qty{974}{\nano\meter}$ is kept constant.

\begin{figure*}[t]
	\centering
	\includegraphics[]{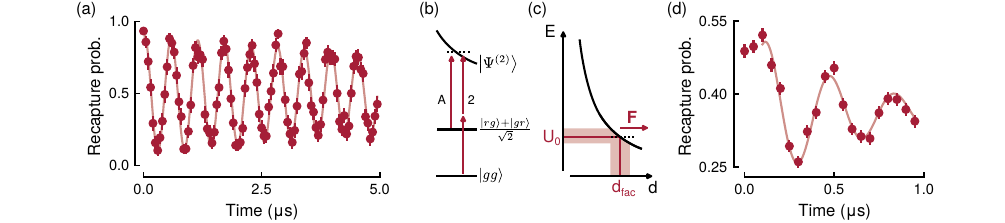}
	\caption{\textbf{Effect of forces on facilitated dynamics of isolated pairs.}
		\textbf{(a)} Reference data showing Rabi oscillations of a single atom in the array.
		To avoid interactions, the Rydberg state was changed to  $45\mathrm{S}$.
		Error bars represent one standard deviation of the mean. We find a $1/e$-coherence time of $\qty{13}{\micro\second}$ at a Rabi frequency of $\Omega=2\pi\times\qty{1.8}{\mega\hertz}$.
		\textbf{(b)} Illustration of the facilitation condition, for which the intermediate state with one atom in the Rydberg state is resonantly coupled to the interaction-shifted pair-Rydberg state $\ket|\Psi^{(2)}>$ (marked by "A").
		The direct second-order coupling (marked by "2") from the pair-ground $\ket|gg>$ to the pair-Rydberg state is also illustrated.
		While the facilitation resonance is shifted to $U_0$, the second-order transition for isolated pairs is found at $U_0/2$.
		\textbf{(c)} Illustration of the coupling between spin and motion in facilitation experiments.
		Positional uncertainty around the facilitation distance $d_\text{fac}$, indicated by the light red area, translates to uncertainty in the interaction strength around $U_0$.
		The gradient of the van der Waals force results in a strong repulsive force $\bm{F}$.
		\textbf{(d)} Coherent oszillations for pairs, initialized in the state $\frac{\ket|rg> + \ket|gr>}{\sqrt{2}}$ and then driven on facilitation resonance.
		Shown is the recapture probability versus coupling time.
		A small recapture probability corresponds to a large probability for the atom to be in the Rydberg state. We find an symmetry-enhanced oscillation frequency of $\Omega_{\mathrm{pair}}=2\pi\times\qty{2.8}{\mega\hertz} \approx 2\pi\times\sqrt{2}\times\qty{2}{\mega\hertz}$ and a $1/e$-decay time of $\qty{0.54}{\micro\second}$.
	}
	\label{fig:si2}
\end{figure*}

We shape both of the laser beams for Rydberg coupling into elliptical beams similar to the light sheet.
The beams are counter-propagating to each other with the $\qty{974}{\nano\meter}$ beam aligned along the light-sheet axis.
Both beams are circularly polarized to maximize the coupling strength to the selected Rydberg state.
For the $\qty{405}{\nano\meter}$ beam we measured a vertical (horizontal) waist of $\qty{35}{\micro\meter}$ ($\qty{300}{\micro\meter}$) and for the $\qty{974}{\nano\meter}$ beam $\qty{10}{\micro\meter}$ ($\qty{170}{\micro\meter}$).
Across the array we measure a root-mean-square variation of $\qty{7}{\percent}$ of the Rabi frequency.
This is dominated by the finite extent of the $\qty{974}{\nano\meter}$ beam in the horizontal direction as well as the vertical Rayleigh range of similar scale.
The variation in Rabi frequency manifests in a large scale harmonic envelope over the atom array.
The small vertical waists of both beams are required to achieve a high Rabi frequency over the entire array, but make the configuration susceptible to small mechanical fluctuations in the experimental setup.
While this effect is reduced for atoms which are at the maximum of the gaussian beams, it is more important for atoms away from the beam center.
We observe this effect as a position dependence of the coherence time in our Rabi oscillations.
This is our main decoherence effect, which, however, has minimal effect on the timescale of $\qty{2}{\micro \second}$ explored in the experiments presented here.
The Rabi oscillations to the  $45\mathrm{S}_{1/2}$ state shown in \autoref{fig:si2}a underline the excellent coherence of single-atom control achieved in our setup.

\section*{Effect of forces}

The main experimental challenge is the strong spin-motion coupling on the facilitation resonance~\cite{zhao2025-cg, schlegel2025-cg, emperauger2025-cga}.
The van der Waals interaction naturally features a strong gradient of $\partial V/\partial d = -6V/d  \approx\qty{10}{\kilo\hertz \per \nano\meter}$ (cf. \autoref{fig:si2}c).
This has two consequences.
First, it translates thermal position disorder into interaction disorder, which hindered early experiments to observe facilitation dynamics~\cite{marcuzzi2017-cg} and requires near-ground-state initial state preparation.
Second, the associated forces are large resulting in an acceleration the order of $10^4 g$ for our parameters, with $g$ the earths gravitational acceleration.
An isolated Rydberg pair features the worst case acceleration per atom, which displaces it by about $\qty{50}{\nano\meter \per \micro\second}$ and thus quickly out of the bandwidth of our transverse field given by $\Omega=2\pi\times\qty{2}{\mega\hertz}$.
In \autoref{fig:si2}d we show that we indeed observe coherent oscillations of Rydberg pairs on the facilitation resonance for more than $\qty{1}{\micro\second}$.
For this measurement, we prepared isolated pairs of $\qty{7}{\micro\meter}$ distance and initialized them with one atom in the Rydberg state by a $\pi$-pulse under Rydberg blockade conditions.
After the pulse we quickly jumped the laser frequency to facilitation resonance.
To remain approximately in the coherent regime, we limited all our studies presented here to similarly short times, even though, clusters are likely less impacted by the motion due to balancing forces~\cite{emperauger2025-cga}.

\section*{Perturbative Rabi dressing of confined clusters}

For the calculation of the shift in the resonance condition for confined and loop-free
clusters of size $k$ caused by the Rabi coupling $\Omega$ we set
$\hbar=1$ and disregard an overall energy shift in the Hamiltonian from the main text,
\begin{equation}
	\hat{H}=U_{0}\sum_{<i,i^{\prime}>}\hat{R}_{i}\hat{R}_{i^{\prime}}+a \cdot U_0 \sum_{\ll i,i^{\prime}\gg}\hat{R}_{i}\hat{R}_{i^{\prime}}-\Delta\sum_{i}\hat{R}_{i}+\frac{\Omega}{2}\sum_{i}\hat{X}_{i}.
\end{equation}
We also include the next-nearest-neighbor (diagonal) interactions of strength $a U_0$ with $a=1/(\sqrt2)^6=1/8$ and disregard any further-range interactions which are much smaller than the $\Omega$ used in the experiment.
As mentioned in the main text, in the classical limit $\Omega\rightarrow0$
the initial state $\ket|\downarrow>^{\otimes N}$ is an eigenstate
at energy $E^{(0)}_{0}=0$, the superscript $(0)$ indicates the classical limit. The eigenenergy $E_{k}^{(0)}$ for a string-like cluster with $k$ excited atoms is $E_{k}^{(0)}=(k-1)U_0-k\Delta+d a U_0$ where the last term depends on the number $d$ of diagonal pairs of excited atoms in the particular cluster shape considered. The resonance condition is determined from  $E^{(0)}_0 \overset{!}= E^{(0)}_k$, it reads
\begin{equation}
	\Delta_{k}^{(0)}=(k-1)U_{0}/k+daU_0/k=\frac{k-1+da}{k}\cdot U_{0}.
\end{equation}
For a linear cluster without diagonal pairs ($d=0$), this is shown as black crosses in Fig.~2b of the main text. However, the experimentally observed resonance conditions are found at significantly \emph{smaller} values of $\Delta$ (larger values of $\widetilde{h}_z$) a trend not explainable by diagonal interactions $d>0$ which shift to larger $\Delta_k^{(0)}$.

We use 2nd order perturbation theory in $\Omega$ to explain the shift
of the resonance condition for a cluster induced by quantum
fluctuations in the form of transverse field (Rabi drive). We will express our results in the following form,
\begin{equation}
	\frac{\Delta_{k}}{U_{0}}=\frac{\Delta_k^{(0)}}{U_0}+O\left(\frac{\Omega^{2}}{U_{0}^{2}}\right).
\end{equation}
\begin{figure}[t]
	\centering
	\includegraphics{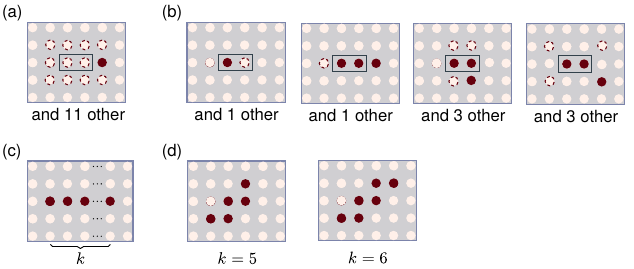}
	\caption{\textbf{Perturbative Rabi dressing of confined clusters.}
	\textbf{(a)} Example for state admixed to the initial state $\ket|\downarrow>^{\otimes N}$ which contributes to $E^{(2)}_0$ in Eq.~\eqref{eq:E(2)_2-E(2)_0}. The rectangle indicates the position of the $k=2$ cluster. The sides marked with a dashed border indicate the other positions of the flipped spin.
	\textbf{(b)} Same, but for $E^{(2)}_2$.
	\textbf{(c)} Linear cluster of $k$ atoms.
	\textbf{(d)} Stair-like clusters of $k=5,6$ atoms.
	}
	\label{fig:si3}
\end{figure}

We start with the smallest possible cluster $k=2$ of two neighboring excited atoms. The resonance condition is defined by a vanishing energy difference $(E_{2}^{(0)}+E_{2}^{(2)})-(E_{0}^{(0)}+E_{0}^{(2)})\overset{!}{=}0$
where superscripts denote orders in $\Omega$. The difference
in the second-order energy correction $E_{2}^{(2)}-E_{0}^{(2)}$
is determined by the number and energies of virtual states which can be reached by a single
Rydberg excitation or de-excitation at the sites of the cluster or nearest-neighbor and next-nearest-neighbor
sites, see \autoref{fig:si3}a,b. Switching processes further away do not affect the energy \emph{difference}.
We find
\begin{equation}
	E_{2}^{(2)}-E_{0}^{(2)}=\left(\frac{\Omega}{2}\right)^{2}\left[\frac{2}{-(-\Delta)}+\frac{2}{-(U_{0}-\Delta)}+\frac{4}{-(U_{0}+aU_{0}-\Delta)}+\frac{4}{-(aU_{0}-\Delta)}\right]-\left(\frac{\Omega}{2}\right)^{2}\left[\frac{12}{-(-\Delta)}\right].\label{eq:E(2)_2-E(2)_0}
\end{equation}
The denominators account for the (classical) energy difference of the virtual cluster states indicated in \autoref{fig:si3}a,b.

In the spirit of perturbation theory we insert the zero-order result
for $\Delta\rightarrow\Delta_{2}^{(0)}=U_{0}/2$ on the right-hand
side. Hence we obtain for the energy difference
\begin{equation}
	0\overset{!}{=}(E_{2}^{(0)}+E_{2}^{(2)})-(E_{0}^{(0)}+E_{0}^{(2)})=
	U_{0}-2\Delta_{2}-2\frac{4a(3a+1)-3}{4a^{2}-1}\cdot\frac{\Omega^{2}}{U_{0}},
\end{equation}
from which we read off
\begin{equation}
	\frac{\Delta_{2}}{U_{0}}=\frac{1}{2}-\frac{3-12a^{2}-4a}{1-4a^{2}}\cdot\frac{\Omega^{2}}{U_{0}^{2}}.\label{eq:Delta2}
\end{equation}
Inserting $\Omega/U_{0}=2.24/11$, $a=0$ and converting $\Delta$ to $\widetilde{h}_{z}$
yields the gray cross in Fig.~2b of the main text which matches the experimental data very well.

For larger clusters with $k>2$ flipped spins we need to generalize the above calculation and the resonance
condition will depend on the cluster geometry. As a geometry-resolved
study is beyond the scope of the present work, we here focus on two
representative cases, (i) the linear cluster and (ii) the stair-like
cluster, see \autoref{fig:si3}c and \autoref{fig:si3}d, respectively.
For the loop-free clusters relevant here, the linear and stair-like clusters are near-extremal cases: the linear cluster is maximally affected by quantum fluctuations and unaffected by interaction shifts beyond nearest neighbors (at the classical level). On the contrary, the stair-like cluster has minimal room for quantum fluctuations and is maximally affected by diagonal interactions. Hence, the results for these two cluster geometries constrain the range in which the non-shape-resolved resonance cluster resonance should fall.

For the linear $k$-cluster, we obtain
\begin{equation}
	\frac{\Delta_{k,lin}}{U_{0}}=\begin{cases}
		1-\frac{1}{k}-\frac{k\left(k^{2}+k+3\right)}{2\left(k^{2}-1\right)}\cdot\frac{\Omega^{2}}{U_{0}^{2}}                                     & :a=0   \\
		1-\frac{1}{k}-\frac{35k^{5}+350k^{4}+58k^{3}-80k^{2}-768k}{14k^{5}+152k^{4}+242k^{3}-664k^{2}-256k+512}\cdot\frac{\Omega^{2}}{U_{0}^{2}} & :a=1/8
	\end{cases}.\label{eq:Delta_k,lin}
\end{equation}

\begin{figure}[t]
	\centering
	\includegraphics[]{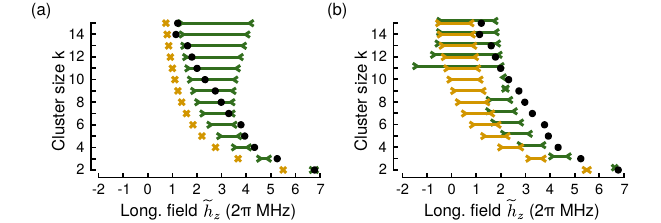}
	\caption{ \textbf{Approximate resonances predicted by a perturbative treatment.}
		\textbf{(a)} Resonance positions without diagonal interactions, $a=0$. In yellow the classical expectation ($\Omega=0$) is presented. Green bars display the range in which the second order perturbative expansion predicts resonances to lie. The edges of the bars are the linear cluster ($<$) and the stair-like cluster ($>$) as representative extreme cases.
		\textbf{(b)} Same as (a) but including diagonal interactions of strength $aU_0=1/8\cdot U_0$. For $k=10$ there is a divergence in the perturbative shift for the stair-like cluster caused by a resonant virtual state.
	}
	\label{fig:si4}
\end{figure}

For a stair-like cluster at $k=3$, an analogous calculation
yields
\begin{equation}
	\frac{\Delta_{3,stair}}{U_{0}}=\frac{a+2}{3}-\frac{9(a(8(a-4)a+5)+16)}{8(a-4)(a-1)(a+2)(2a+1)}\cdot\frac{\Omega^{2}}{U_{0}^{2}}.
\end{equation}
For $a=0$, the range spanned by $\Delta_{3,stair}$ and $\Delta_{3,lin}$ is shown in Fig.~2b of the main text, the value for the linear cluster matches the (non-shaped resolved) experimental data point reasonably well.

For the stair-like cluster with $k>3$ (even and odd $k$), we obtain similarly
\begin{equation}
	\frac{\Delta_{k>3,stair}}{U_{0}}=\begin{cases}
		1-\frac{1}{k}-\frac{3k}{2(k-1)}\cdot\frac{\Omega^{2}}{U_{0}^{2}}                                                                                                                & :a=0   \\
		\frac{9}{8}-\frac{5}{4k}-\frac{5472k^{5}-73610k^{4}-74800k^{3}+102000k^{2}+150000k}{6480k^{5}-64800k^{4}-18125k^{3}+181250k^{2}+12500k-125000}\cdot\frac{\Omega^{2}}{U_{0}^{2}} & :a=1/8
	\end{cases}.\label{eq:Delta_k,stair}
\end{equation}
Fig.~\ref{fig:si4} shows the ranges of the above theoretical $\Delta_k$ bounded by the linear and stair-like cluster together with the experimental data reproduced from the main text. Yellow markers show the classical expectations while green markers indicate Eqns.~\eqref{eq:Delta2}-\eqref{eq:Delta_k,stair}. While panel (a) uses only nearest-neighbor interactions $a=0$, panel (b) employs also the diagonal next-nearest-neighbor coupling, $a=1/8$. Note that in this case also the classical cluster energies are shape-dependent (yellow ranges).

\bibliography{bibliography}

\end{document}